%% file: ms_astroph.tex
\documentclass[12pt,preprint]{aastex}

\newcommand{\htwo}{H\,{\sc ii}}
\usepackage{epsfig}
\usepackage{color}
\begin{document}

\title{Spectroscopic Assessment of {\it WISE}-based Young Stellar Object Selection}

\author{Xavier Koenig\altaffilmark{1}, Lynne
  A. Hillenbrand\altaffilmark{2}, Deborah L. Padgett\altaffilmark{3},
  Daniel DeFelippis\altaffilmark{2}}

\altaffiltext{1}{Department of Astronomy, Yale University, New Haven,
  CT 06511, USA} \altaffiltext{2}{Department of Astronomy, California
  Institute of Technology, Pasadena, CA 91125, USA}
\altaffiltext{3}{NASA Goddard Space Flight Center, Greenbelt, MD
  20771, USA}

\begin{abstract}
We have conducted a sensitive search down to the hydrogen burning
limit for unextincted stars over $\sim$200 square degrees around
Lambda Orionis and 20 square degrees around Sigma Orionis using the
methodology of \citet{koenig14}. From {\it WISE} and 2MASS data we
identify 544 and 418 candidate YSOs in the vicinity of Lambda and
Sigma respectively. Based on our followup spectroscopy for some
candidates and the existing literature for others, we found that
$\sim$80\% of the K14-selected candidates are probable or likely
members of the Orion star forming region. The yield from the
photometric selection criteria shows that {\it WISE} sources with $K_S
-w3 > 1.5$ mag and $K_S $ between 10--12 mag are most likely to show
spectroscopic signs of youth, while {\it WISE} sources with $K_S -w3 >
4$ mag and $K_S > 12$ were often AGNs when followed up
spectroscopically. The population of candidate YSOs traces known areas
of active star formation, with a few new `hot spots' of activity near
Lynds 1588 and 1589 and a more dispersed population of YSOs in the
northern half of the \htwo\ region bubble around $\sigma$ and
$\epsilon$ Ori. A minimal spanning tree analysis of the two regions to
identify stellar groupings finds that roughly two-thirds of the YSO
candidates in each region belong to groups of 5 or more members. The
population of stars selected by {\it WISE} outside the MST groupings
also contains spectroscopically verified YSOs, with a local stellar
density as low as 0.5 stars per square degree.
\end{abstract}


\section{Introduction}
While the data return from the {\it Spitzer Space Telescope} vastly
improved our picture of star formation and circumstellar disk
evolution through study of a large number of molecular clouds at
mid-infrared wavelengths, {\it Spitzer} was a pointed mission with a
small field of view. The more recent {\it WISE} mission mapped {\it
  the entire sky} in 3.4, 4.6, 12, and 22 $\micron$ filters to a
specified 5$\sigma$ point source sensitivity of 0.08, 0.11, 1, and 6
mJy (or 16.5, 15.5, 11.3, and 7.9 mag) at spatial resolution of
6--12$\arcsec$. With sky coverage better than IRAS and spatial
resolution within a factor 2--3 as good as {\it Spitzer}, a clear
advantage of {\it WISE} is its coverage of many regions of interest
that {\it Spitzer} simply missed. Sensitivity with {\it WISE} is
comparable at all but the 22 / 24 $\micron$ bands to {\it Spitzer's}
shallow surveys, e.g. GLIMPSE in the Galactic plane.

Circumstellar disks surround young stars for the first several Myr of
their lives. These disks create infrared through millimeter
wavelength excesses due to thermal emission from dust that is
distributed over distances from a few hundredths to several hundred AU
from the star. {\it Spitzer} and {\it WISE} both probe the infrared
excess emission that arises within a few AU.

As {\it Spitzer} mapped primarily the young constituents of known
molecular clouds, our knowledge of disk evolution time scales based on
{\it Spitzer} studies is still biased and somewhat limited. {\it
  WISE}, however allows us to fill in many gaps in studies of nearby
star forming regions. Particularly noteworthy wide-field studies that
were enabled by {\it WISE} include those by Rebull et al. (2011) on
Taurus, Rizzuto et al. (2012) and Luhman \& Mamajek (2012) on Upper
Sco, and Koenig et al. (2012) and Koenig \& Leisawitz (2014) on outer
Galaxy regions.

The Orion molecular cloud and young star complex is a benchmark star
forming region (see Bally 2010 for a review). Orion's prominence is
due to its proximity ($\sim$400~pc) as well as its representation of
both high and low mass star formation, of clustered and isolated star
formation, and of future, ongoing, and recent star formation. The
$\sigma$ Ori cluster, just west of the Horsehead Nebula and below the
easternmost ``belt star" in the constellation Orion, is a 3--5 Myr old
cluster associated with the central O9.5V star. Its stellar population
has been studied by Wolk (1996), \citet{sherry}, \citet{caballero},
and \citet{lodieu} and currently contains $\sim$350 confirmed members
and another $\sim$300 photometric candidates. There is relatively
little molecular gas in the region \citep[e.g.][]{lang}; it is located
between the Orion A and Orion B giant molecular clouds. The $\lambda$
Ori cluster, near the famed Betelguese in Orion, is a 4--8 Myr old
cluster associated with the central O8 III star and contains ten
additional B stars. Its low mass population has been studied by Dolan
\& Mathieu (1999, 2001, 2002), \citet{barrado}, and \citet{sacco}
though membership is vastly incomplete and knowledge of spectral types
limited.  A surrounding ring includes dust and swept-up neutral and
molecular gas \citep[e.g.][]{lang} and potentially younger stars than
those immediately surrounding $\lambda$ Ori. There is also suggestion
of a previous supernova explosion \citep{cunha}. For reviews of these
two regions, see \citet{walter2008} and \citet{briceno2008}
respectively, for the $\sigma$ Ori cluster and the extended Orion
OB1a/OB1b region included in the presentation below, and
\citet{mathieu2008} for the $\lambda$ Ori region.

Rich clusters in the age range of $\sigma$ Ori and $\lambda$ Ori are
relatively rare. Complete surveys with {\it WISE} of these two
clusters are thus important for census building, and for overall and
uniform studies of circumstellar disk properties and for our
understanding of disk evolution.
Analysis of 2MASS data alone shows that these two clusters indeed have
a lower fraction of {\it near-infrared} excess relative to many 1--3
Myr old clusters still assocatiated with molecular gas.  A key
question is whether the {\it mid-infrared} excess fraction is also
proportionally lower or similar; the former would imply disk depletion
at all radii relatively quickly whereas the latter would suggest
inside-out depletion.

\section{The {\it WISE} view of $\sigma$ and $\lambda$ Orionis}

\begin{figure}
\centering
\includegraphics[width=8.5cm]{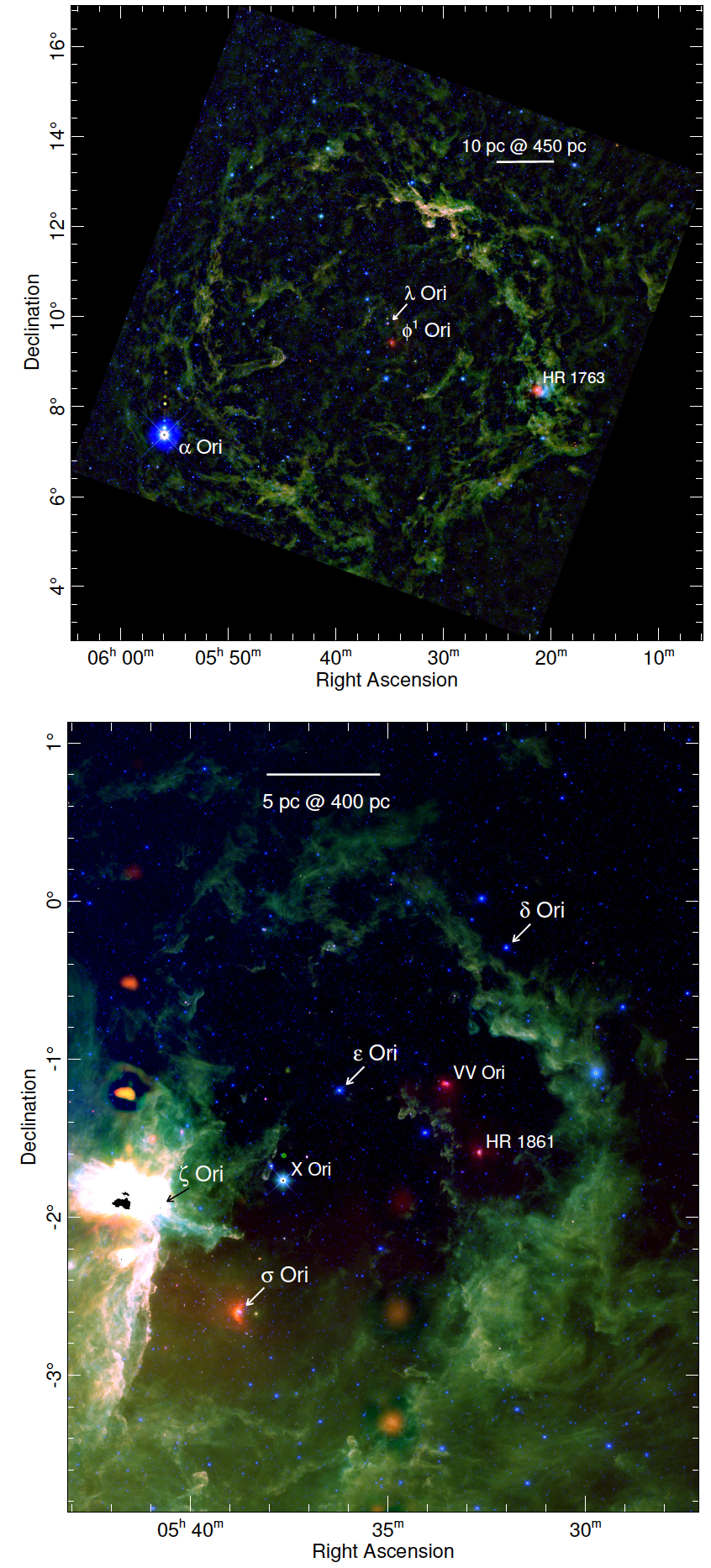}
\caption{{\it WISE} 3-color image mosaics for the $\sigma$ (upper
  panel) and $\lambda$ Orionis (right panel) regions. The image assigns
  the 3.4, 12, and 22~$\micron$ channels to blue, green and red in the
  image respectively. The large, diffuse red spots appearing at
  regular intervals northward from NGC~2023/2024 and from $6^h35^m$,
  $-4\degr$ are bright source image latents from
  $w4$.~\label{fig:3color}}
\end{figure}

Figure~\ref{fig:3color} shows a 3-color composite of {\it WISE} image
data in the 3.4, 12, and 22~$\micron$ ($w1, w3, w4$) bands for the
$\sigma$ and $\lambda$ Orionis fields. Emission at 12~$\micron$ (green
in the figure) is dominated by bright PAH emission that traces cloud
surfaces and reveals the well known $\lambda$ Orionis ring, as well as
the ring around the extended $\sigma$ Orionis field. The 3.4~$\micron$
band (assigned to the blue channel in the figure) has some
contribution from PAH emission lines, but mainly picks out stellar
photospheric emission from foreground and background stars. The
22~$\micron$ band (in red) captures thermal emission from dust grains
which highlights cloud surfaces in the same way as $w3$, but also
heated dust close to OB stars. A SIMBAD search shows that in the
$\lambda$ Ori region this feature is seen around $\phi^{1}$ Orionis
(B0.5III) and HR~1763 (B1V). In the $\sigma$ Ori region a bow-wave
feature around $\sigma$ Orionis, \citep[see][]{ochsen14} and halos
around VV~Ori (B1V) and HR~1861 (B1IV) are picked out by bright,
extended 22~$\micron$ emission. These stars are labeled in the figure,
as well as the three main belt stars of the Orion constellation and
X~Ori, a Mira variable evolved star.

Note that the other red, evenly spaced, diffuse patches of emission
aligned vertically with the bright emission in the eastern part of the
$\sigma$ Ori region and also appearing in the lower, central part of
the panel are image artifacts in the 22~$\micron$ {\it WISE} images.

\section{Initial YSO Selection Criteria}

The combination of 1.2, 1.6, and 2.2 $\micron$ photometry from 2MASS
with the 3.4, 4.6, 12, and 22 $\micron$ photometry from {\it WISE} is
a sufficient lever arm on the stellar (e.g. from $J-K$ colors) to
circumstellar (e.g. from $w1-w2$ and $w1-w4$ colors) regime that
robust identification of evolving and dissipating young circumstellar
disks can be made.

We initially extracted sources from the regions surrounding $\sigma$
and $\lambda$ Orionis from the {\it WISE} All Sky point source
catalog. For the former we extracted a region bounded by
84.0$<$RA$<$85.5 degrees and $-3.0<$Dec$<2.05$ degrees (J2000.0), for
the latter a polygon with vertices at (RA, Dec) = (77.1 11.6; 87.7
15.4; 90.8 6.5; 80.5 2.8) to encompass the full extent of the large
ring of emission that surrounds the $\lambda$ Orionis \htwo\ region,
as seen in the {\it WISE} band 3 atlas images at 12~$\micron$.

The {\it WISE} All Sky catalog also provides photometry in
the near-infrared $J$, $H$ and $K_S$ bands from 2MASS. The {\it WISE}
All Sky photometry pipeline uses a match radius of 3$\arcsec$ to
identify the 2MASS counterparts of {\it WISE} sources.

We then applied an early version of the {\it WISE-2MASS} YSO
classification and galaxy filtration scheme of \citet{koenig12}. As
described by Koenig et~al. the scheme first applies {\it WISE} color
and magnitude cuts to remove background unresolved star-forming
galaxies and active galactic nuclei (AGNs). These objects are
typically redder and fainter than the majority of young stars in
Galactic star-forming regions. A further round of cuts is made to
remove objects likely to be shocked blobs of gas in young star
outflows (bright at {\it WISE} band 2) and spurious detections of
bright nebular emission (objects with very red $w2-w3$ color). The
remaining sample is then classified into young star `classes' using
{\it WISE} colors, matching the categories laid out by
\citet{greene94} as best as possible. This scheme does not attempt to
find either Class III or Flat SED objects, however, only Class II and
Class I. Additional young stars are selected based on combined {\it
  WISE} and 2MASS colors, aiming to retrieve those objects missed due
to their {\it WISE} band 3 detection being obscured by bright nebular
background emission. The final steps of the scheme look for candidate
transition disk objects \citep{strom89} with red $w2-w4$ colors and
test whether the previously classified Class I objects have red enough
$w2-w4$ colors, or are more likely to be the more evolved Class II.

In the initial selection process for this study, we differed from the
\citet{koenig12} scheme just described in the follow ways.  We did not
set an upper limit to the magnitude error in a given band, but simply
required that if photometry was needed for a given classification
criterion, that the photometric uncertainty should have a non-null
value. We also used a more lenient $w2$ magnitude cut for
star-forming/PAH galaxies of $w2<13.5$ and we did not apply the very
last cut of the scheme that removes objects at the blue end of the
{\it WISE} T Tauri star locus. The region of {\it WISE} color-color
space roughly defined by $0.5<w2-w3<1.1$ and $0.1<w1-w2<0.4$ is
occupied by older, more evolved Classical Be stars \citep[as
  subsequently shown by][hereafter, K14]{koenig14}, but it can also
contain young `transition disk' objects
\citep[see][]{cieza12}. Contamination of {\it WISE} band 4 at
22~$\micron$ by fake detections due to nebulosity produced a large
number of candidate transition disks in this initial YSO candidate
list. We thus required a signal to noise at band 4 of at least 4.5,
and set an upper limit to the $w2-w4$ of 6 to suppress falsely
classified objects.

Having produced a list of candidate YSOs for both $\sigma$ Ori and
$\lambda$ Ori, we additionally constrained the source lists before
each spectroscopic run at the telescope, as described below.

\section{Spectroscopic Target Selection and Observations}

In total, 230 sources were observed spectroscopically at the Palomar
Observatory 200" telescope. All data were collected with the Double
Spectrograph (originally comissioned by Oke \& Gunn 1982 but with many
upgrades since that time) using a dichroic at 4800 A and a 1200 l/mm
red and a 300 l/mm blue grating.  Exposures of FeAr and HeNeAr lamps
were obtained for the blue and red side wavelength calibration.

The spectra were collected over several years, with evolving selection
criteria.

\subsection{2009 Spectra}

As part of an unrelated program studying low mass stars in $\sigma$
Ori \citep{cody14}, spectroscopic observations were taken in 2009 of
known members that were later revealed to be of interest from the
vantage of the current {\it WISE} study. These objects were selected
for spectroscopy based on their presence in the fields photometrically
monitored by \citet{cody10} and lack of a published spectral type at
that time.

As discussed by \citet{cody12}, observations took place on the nights
of 18--21 January and 19--20 December with 68 unique low mass stellar
and brown dwarf objects in $\sigma$ Ori observed.

\subsection{2012 Spectra}

For the first run dedicated to {\it WISE} follow-up in $\lambda$ Ori
and $\sigma$ Ori, we required all stars to have non-null photometric
error in {\it WISE} bands 1 and 4 and to not suffer from the
diffraction spike, scattered light halo, optical ghost or latent
contaminant flags (SQL query: {\it cc\_flags} not matches
'[DHOP]'). To minimize obvious contamination by background galaxies,
we required the 2MASS Extended Source Catalog proximity column {\it
  xscprox} to be either null or $>30\arcsec$. Finally, using the {\it
  WISE} band 4 image atlas we inspected all the YSO candidates and
rejected objects that were extended or whose 22~$\micron$ emission
appeared to be offset from the shorter wavelength centroids by more
than 2$\arcsec$. In both fields we noticed a concentration of objects
with red colors that appears at magnitudes fainter than $J=16$ and
$K_S=14$. The spatial distribution of these detections is roughly
uniform; thus they are likely mostly extragalactic, so we also cut
them from further consideration. We obtained optical photometry for
all sources where available from the
NOMAD\footnote{http://www.usno.navy.mil/USNO/astrometry/optical-IR-prod/nomad}
online database and restricted our targets to objects with $R$
magnitude less than 18.3. For $\sigma$~Ori we were able to use imaging
data from the Sloan Digital Sky Survey\footnote{www.sdss.org} to
confirm that objects fainter than $J=16$ are indeed dominated by
galaxies.

In $\sigma$~Ori we obtained 29 objects with DBSP on
01 January, 2012, UT, selecting those YSO candidates with $K_S < 14$
and without existing spectral types in the literature or in Cody's
thesis work discussed above.  Most objects having $J < 11$ and $w1-w2
> 0.1$ were observed, as were most objects at all brightnesses with
$w1-w2 > 0.7$. Among objects with $J > 11$, the observations focused
on those without previous designations in SIMBAD, i.e. {\it WISE} is
the first that attention has been called to them. A few objects
meeting the above criteria had roughly equal brightness, close
companions within a few arcsec in 2MASS or on the telescope guide
camera when slewing to the field. These were excluded from observation
due to concerns about contamination in the excess selection, since we
didn't have enough information to attribute $w3$ or $w4$ excesses
between close equal sources.

In $\lambda$~Ori we obtained 40 objects with DBSP on
02 January, 2012, UT, selecting YSO candidates with $J < 16$ and $w1 <
14$ and a $K_S < 14$ cut during the observing. We also limited
ourselves to targets south of Declination +09 so as to avoid the areas
previously studied by others, namely, the 1 deg$^2$ around
$\lambda$~Ori itself and the B30 and B35 regions, with the goal being
that our work with {\it WISE} is unique in the lower Declination
areas.

The prioritization for observations was a combination of brightness
(brighter objects at high airmass) and $w1-w2$ color. Nothing with
$w1-w2 < 0.15$ was observed (i.e., none of the Class III/transition
disk candidates). We focused on the reddest $w1-w2$ colors. At the
faint end, some spectra turned out to be active galactic nuclei
(AGNs), but in general we attempted to observe the faintest YSO
candidates. Spatially, the clustering of excess objects coincident
with the $\lambda$~Ori ring towards the southeast was prioritized;
this spatial cut also captured many of the reddest $w1-w2$ objects
that were not so faint as to be likely extragalactic contaminants.  In
examining regions to the west of $\lambda$~Ori, almost every {\it
  WISE}-selected YSO candidate with $J < 12.5$ was observed.

\subsection{2013 Spectra}

In 2013, all targets were in the $\lambda$~Ori field, again with a
focus on the western and southern parts of the region.  We selected
YSO candidates with $8 < w1 < 13$ to avoid both infrared-bright
asymptotic giant branch stars (AGBs) and faint AGNs, although we did
not impose a color criterion. In total, 80 sources were observed on 2,
4, 5 February 2013, UT.

For this run, we did not apply the additional {\it WISE} band 4
quality assessment, or the requirement on the {\it cc\_flags}
parameter that were used for the 2012 run. One result potentially
attributable to this modification is that the fraction of observed
sources with emission lines was lower than in the earlier spectra.

\section{Spectroscopic Data Reduction and Analysis}

Two-dimensional CCD images from both the blue and red channels of the
spectrograph were corrected for detector bias and flat field and
one-dimensional spectra were extracted, wavelength calibrated, and
flux calibrated, all within the {\it twodspec} and {\it onedspec}
packages of IRAF. In 2012, Feige 34, Feige 110, Gl 38-31, HD 19445,
and HD 199178 were used as flux standards while in 2013, Feige 34, HD
93521, Hiltner 600, and G 191B2B served the purpose. For our setup,
the full spectral range on the blue side was $\sim$3525--5075\AA, and
on the red side 4660--10,940\AA, though mediated by the dichroic and
the atmosphere.

The stars were classified visually by comparison to the Jacoby, Hunter
\& Christian (1984) and the Allen \& Strom (1995) standards. The
DoubleSpec blue side and red side spectra were classified
independently, with the final spectral type taken as the intersection
of the ranges concluded on each side.

Emission and absorption lines were studied in each object. Equivalent
widths were measured for H$\alpha$, \ion{Li}{1} 6707 \AA, the
\ion{Ca}{2} triplet at 8498, 8542 and 8662 \AA, and related H and K
lines at 3933 and 3968 \AA. Other emission and absorption lines noted
in some spectra included [\ion{O}{1}] 6300, 6363 \AA, [\ion{S}{2}]
6717, 6732 \AA, \ion{O}{1} 8446 \AA, \ion{Na}{1} 5890 \AA, \ion{He}{1}
5876 \AA, and the Balmer series of Hydrogen out to H$\delta$. Several
objects with the Balmer jump clearly in emission were also identified.

We found that objects with $K_S -w3 > 1.5$ mag and $K_S $ between
10--12 mag were most likely to show spectroscopic signs of youth such
as strong H$\alpha$ or the \ion{Ca}{2} triplet in emission. Extremely
red objects with $K_S -w3 > 4$ mag and $K_S > 12$ were more likely to
be AGNs.

\section{Updated YSO Selection Criteria}
The initial {\it WISE}-based selection criteria and spectroscopic
follow up observations described above served to obtain a sample of
young stars with optical spectra in the $\lambda$ and $\sigma$ Orionis
regions. Subsequent work by K14 showed that this original YSO finding
scheme generates a large number of spurious candidate young stars.  We
thus implemented for final presentation here the more robust YSO
finding procedure detailed in K14. The K14 scheme begins in the same
way as K12 by removing extragalactic contaminants (AGNs and
star-forming galaxies).  It does not include a step to remove
detection of sources that appear red due to nebular emission because
it relies on the fact that YSO candidates occupy loci in color space
that naturally separate them from these detections. As with K12, the
scheme then uses {\it WISE} and then {\it WISE}+2MASS color criteria
to identify and classify Class I and II YSOs. Final steps select
candidate transition disk objects with red $w3-w4$ colors and retrieve
candidate protostars from objects flagged as AGNs that are bright at
$w4$. A new feature of the K14 scheme is the addition of an explicit
phase of AGB star removal that uses cuts in the {\it WISE} $w1-w2$
vs. $w3-w4$ color-color and $w1$ vs. $w1-w2$ color-magnitude
diagrams. Table~\ref{tab:k14class} presents a comparison of the number
of YSO candidates produced by the K12 method described in Section 3
with the number produced by the K14 scheme, over the area of sky shown
in Figure~\ref{fig:3color}. The newer selection algorithm clearly
finds a greatly reduced number of candidates as a consequence of its
more conservative design.

\input{tab1.tex}

\begin{figure}
\centering
\includegraphics[width=8.5cm]{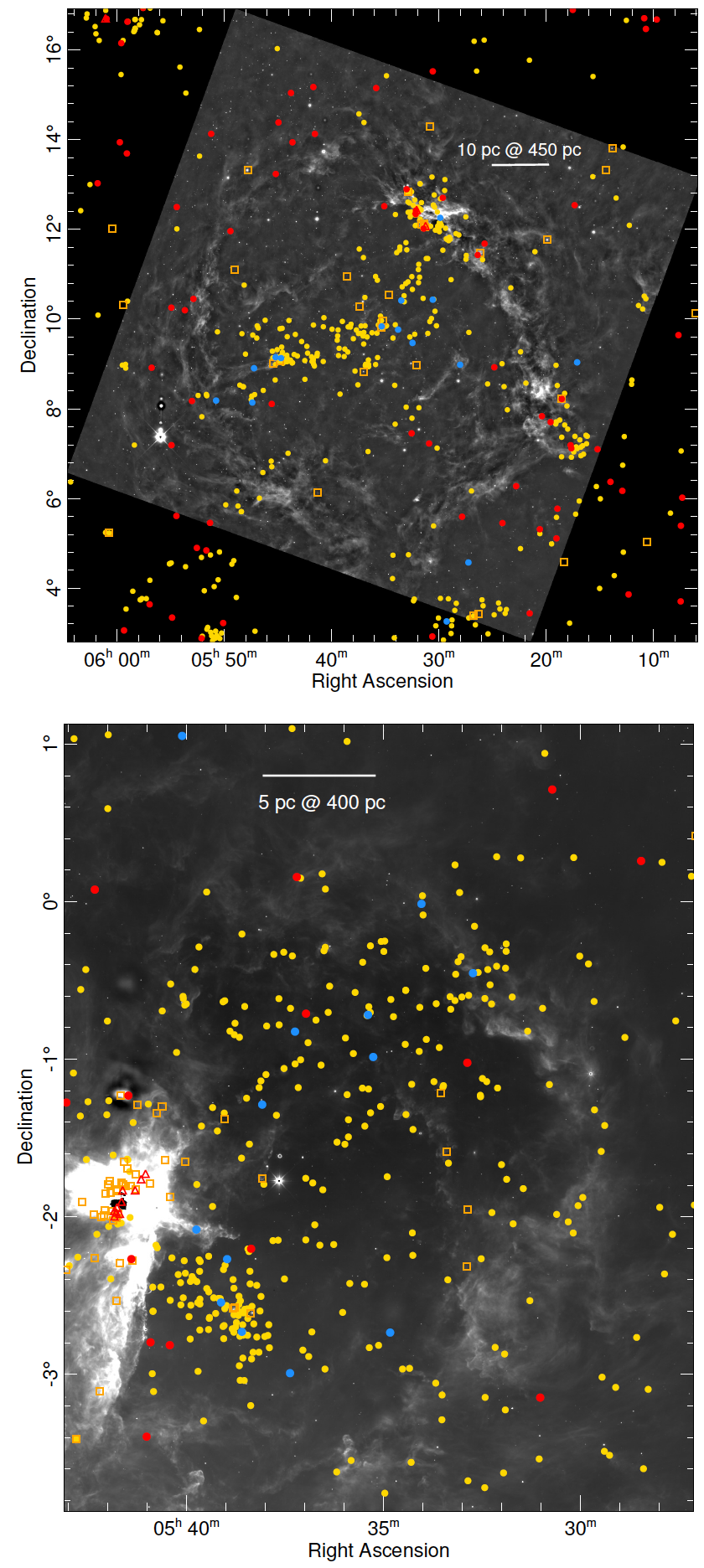}
\caption{Distribution of YSO candidates in $\sigma$ and $\lambda$
  Orionis (upper and lower panels respectively) as identified by the
  scheme of Koenig \& Leisawitz (2014). Red points and triangles:
  candidate Class I's from {\it WISE} or {\it WISE}+2MASS. Yellow
  points and squares: candidate Class II's from {\it WISE} or {\it
    WISE}+2MASS. Blue points: candidate transition disk
  objects.~\label{yso-dist}}
\end{figure}

The spatial distribution of YSO candidates found using the K14 scheme
in $\lambda$ and $\sigma$ Orionis is shown in
Figure~\ref{yso-dist}. To assess the new contribution made by our
survey to the distribution of YSO candidates in the two regions, we
carried out a census of the relevant literature. Both the $\lambda$
and $\sigma$ Orionis regions have been surveyed by many authors. The
main studies in $\lambda$ Ori are those by \citet{hernandez10} and
\citet{dolan02}. In $\sigma$ Ori we compile the Mayrit catalog of
\citet{mayrit} and the studies by \citet{penram}, \citet{briceno},
\citet{megeath}, \citet{bejar} and \citet{luhman}.

\begin{figure}
\centering
\includegraphics[width=8.5cm]{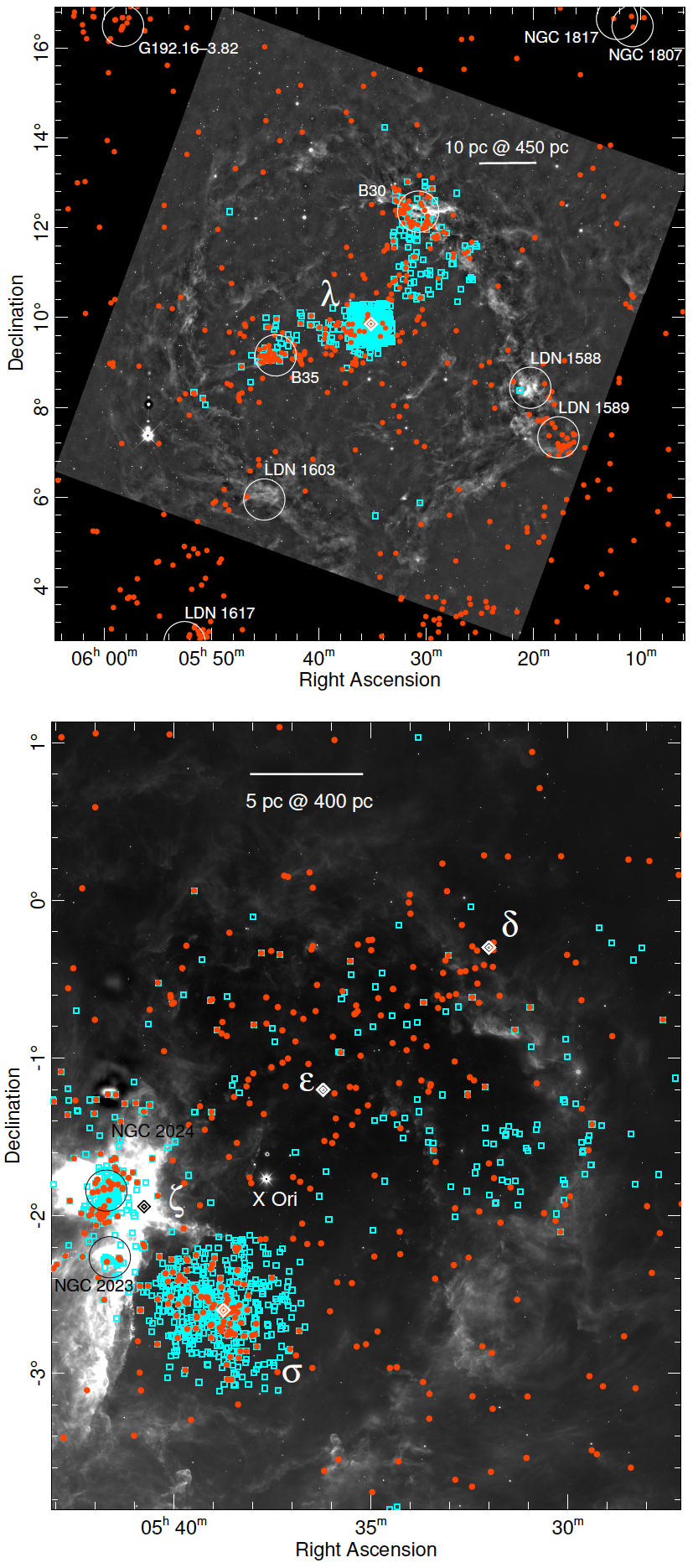}
\caption{Distribution of literature YSOs from the papers cited in the
  text (cyan box points) and {\it WISE} YSO candidates (red-orange dot
  points). $\lambda$, $\epsilon$ and $\sigma$ Orionis are plotted as
  white diamonds. The locations of other notable dark clouds or
  clusters are also marked.~\label{lit-dist}}
\end{figure}

\begin{figure}
\centering
\includegraphics[width=8.5cm]{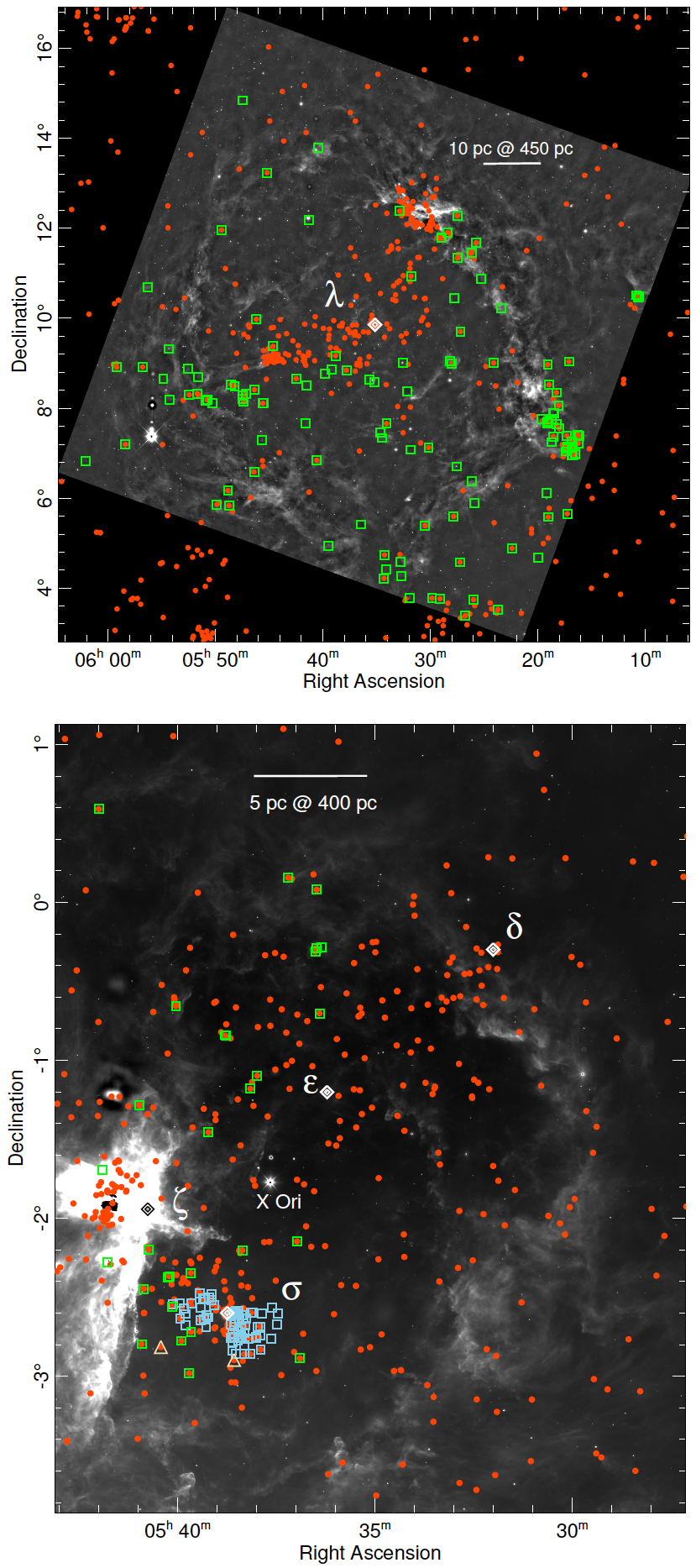}
\caption{Distribution of spectra acquired in this paper (green box
  points), the Cody et~al. (2010) sample (blue box points) and two
  YSOs from Riaz et~al. (2015) (light triangle points). YSO candidates
  are shown as red orange dot points. $\lambda$ Ori, and in the lower
  panel $\epsilon$, $\delta$ and $\sigma$ Orionis are plotted as white
  diamonds and $\zeta$~Ori as a black diamond.~\label{spec-dist}}
\end{figure}

Figure~\ref{lit-dist} shows how the compiled list of previously known
likely members of the two regions relate to the newly identified {\it
  WISE} YSO candidates found in this paper. Figure~\ref{spec-dist}
shows the distribution of the targets followed up with optical
spectroscopy in the two regions. The upper panels of
Figures~\ref{yso-dist}, \ref{lit-dist} and \ref{spec-dist} show that
our extracted sample of YSO candidates extends beyond the boundaries
of the $\lambda$ Ori ring. G192.16$-$3.82 is an \htwo\ region at a
distance of 1.52~kpc \citep{shiozaki}, with an associated young
cluster \citep{carpenter93}. LDN~1617 is known for its association
with Herbig Haro outflow sources and is at a distance $\sim$400~pc
\citep{kajdic}.

\subsection{YSO Spectral Slopes}

While the scheme of K14 provides an assessment of the distribution and
a simple breakdown of the evolutionary state of the YSO candidates
that it selects, a more commonly used means of dividing young objects
with infrared excess emission is to use the slope of their spectral
energy distributions (SEDs) in the near and mid-infrared. We follow
the methodology of \citet{greene94} developed for the availability of
K-band and 25 $\mu$m IRAS measurements, and compute $\alpha$, the
slope of the SED, for each YSO candidate over the longest possible
wavelength baseline available from 2MASS and {\it WISE} (note: we do
not perform a fit to the SEDs). Ideally we compute $\alpha$ from the
slope between the 2MASS $K_S$ band to 22 $\mu$m at {\it WISE} band 4,
requiring non-null $K_S$ band photometric error, signal to noise in
$w4 \geq 5$ and $0.3<\chi^2_{w4}<1.7$. If a source does not meet the
$K_S$ band requirement we derive $\alpha$ from the slope between $w1$
and $w4$, requiring signal to noise in $w1 \geq 3$. If the band 4
photometry fails the above criteria we use {\it WISE} band 3 with the
same test on signal to noise and $\chi^2$ at $w3$, paired with either
2MASS $K_S$ or $w1$ as appropriate. Finally we resort to the slope
between 2MASS $H$ and {\it WISE} $w2$ bands if $K_S$, $w3$ and $w4$
all fail the above tests. $\alpha$ is computed from the following
equation:

\begin{equation}
\alpha = \frac{\mathrm{d} \log \lambda F_\lambda}{\mathrm{d} \log \lambda}
\end{equation}

Note that for all objects, we attempt to remove the effects of dust
extinction by looking up the value of $A_K$ in the nearest pixel in
the $K$-band extinction map of \citet{lombardi11}. We compute the
extinction in the 2MASS bands using the extinction law of
\citet{rieke85} and in the {\it WISE} bands using the interpolation
presented in \citet{koenig14}. Having derived a value of $\alpha$ for
each YSO candidate, we divide them into the categories of
\citet{greene94} as Class I ($\alpha \geq$ 0.3), Flat ($0.3>\alpha$
$\geq -0.3$), Class II ($-0.3>\alpha \geq -1.6$) and Class III
($\alpha < -1.6$). Note that the Class III category does not signify
`no infrared excess,' but rather `weak excess.'  We tabulate the
resulting breakdown of source types in Table~\ref{tab:k14class}. In
both fields, some fraction of the K14 Class II sources are categorized
as Class III in the $\alpha$ prescription, and some are placed in the
`Flat-SED' class, although no more than 20\% in either case. However,
from the K14 Class I sources, 63\% of the candidates in $\sigma$ Ori
move to the Flat SED class under the $\alpha$ calculation, compared
with 25\% in $\lambda$ Ori. In general this suggests that the Class I
sources in $\lambda$ Ori extend to much redder colors such that their
slopes $\alpha$ are consistent with a Class I source in that
prescription. However, it should be noted that the most embedded
clusters in the $\sigma$ Ori region, those near to NGC~2023 and 2024,
are poorly characterized in this survey, owing to the high degree of
saturation in the {\it WISE} band 3 and 4 images in that area. This
issue limits the robustness of the comparison of the $\alpha$ SED
class distributions between $\lambda$ and $\sigma$ Ori.

The Class III category in the $\alpha$ prescription is an overlapping
set with K14's Class II and transition disk sources, because its only
definition is a steeply negative SED slope. Since transition disk
candidates are selected on the basis of weak excess in $w1-w2$ color
and strong excess in $w3-w4$, most of the transition disk objects in
our samples are actually counted as Class II YSOs by the $\alpha$
prescription (14/15 from the $\lambda$ Ori sample, 11/13 from $\sigma$
Ori).

\section{Assessment of the WISE-Selected YSO Candidates}

\subsection{Spectroscopic Success Rate in YSO Confirmation}
In $\lambda$ Orionis, 75 of our spectra were YSO candidates from the
K14 selection scheme. In $\sigma$ Orionis, we obtained spectra of 48
K14 YSO candidates. We examined how many of these targets showed
emission in either the H$\alpha$ (equivalent width $<-10$~\AA) or
\ion{Ca}{2} 8542 (equivalent width $< 0$~\AA) lines as a simple test
of `youth,' although it should be noted that many M stars show
H$\alpha$ in emission. In $\lambda$ Orionis, 50/75 objects (67\%) met
these criteria, while in $\sigma$ Orionis 28/48 spectra (58\%) met
these requirements.  While a more general indicator of youth is
provided by the Lithium \ion{Li}{1} 6707 \AA\ line, our Palomar
spectroscopic data are not only too low in resolution for this to be a
reliable youth test, but were also found to suffer significant
uncertainty due to a feature at this wavelength in the flat field that
can be traced to the long decay time of a dome light, resulting in
additional absorption at this wavelength in the final spectra.  For
this paper, we choose not to use the measured \ion{Li}{1} equivalent
width measurements.

\subsection{Spectral and Mid-Infrared Properties}
\begin{figure}
  \centering
  \includegraphics[width=16cm]{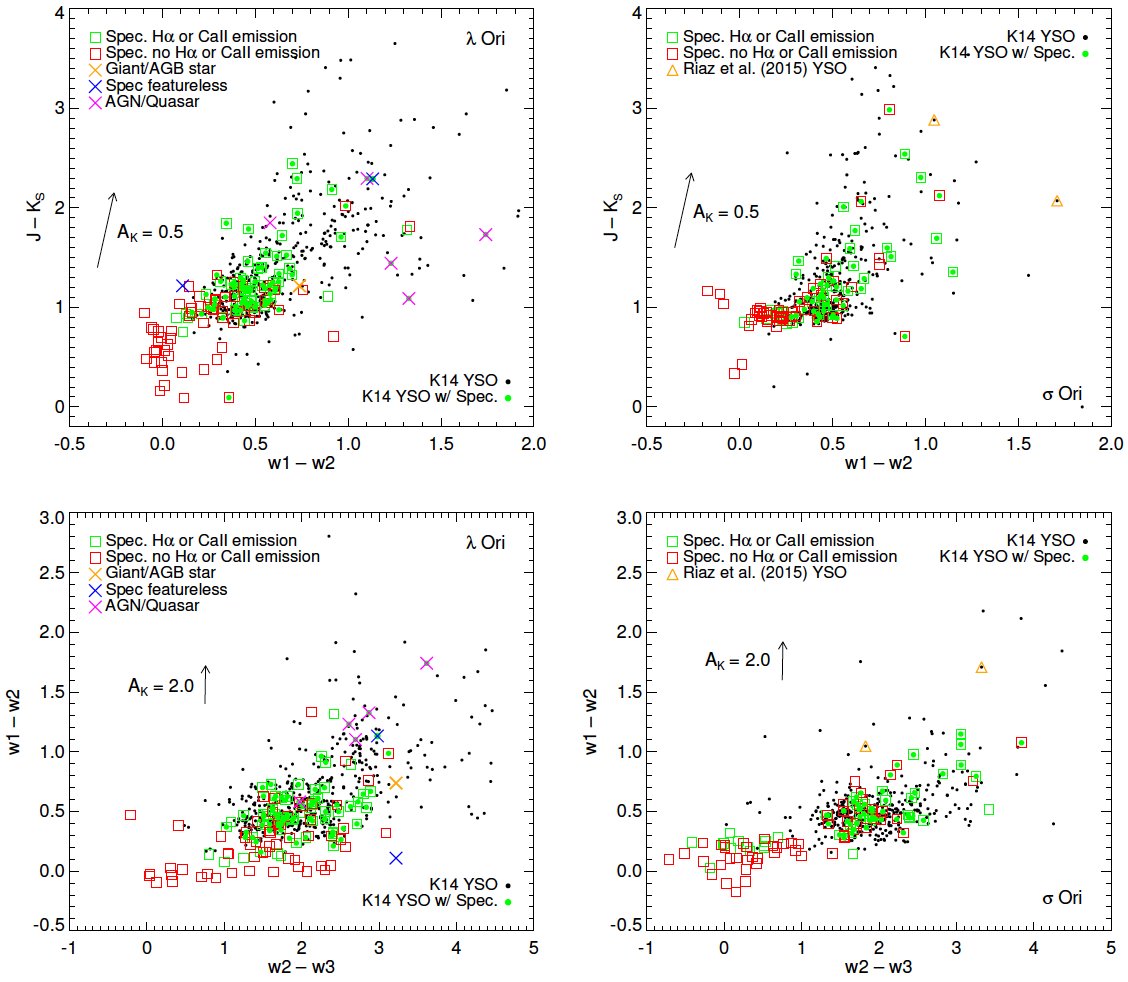}
  \caption{Infrared color-magnitude diagrams of the K14-selected YSO
    candidates in $\lambda$ and $\sigma$ Orionis (black points, or
    green points if a spectrum was acquired). Additional plot symbols
    show spectroscopic targets, colored as in figure
    legend.~\label{fig:color-color}}
\end{figure}

In Figure~\ref{fig:color-color} we show color-color diagrams of the
K14 {\it WISE} YSO candidates in the two Orion regions. We overlay the
colors of the spectroscopic sample acquired in each field, separating
those with and without H$\alpha$ or \ion{Ca}{2} 8542~\AA\ above our
threshold values, and distinguish those spectra that correspond to K14
YSO candidates and those that are not picked out by that selection
process. In the panel for $\sigma$ Ori, we also show the location of
the two objects investigated by \citet{riaz}, both of which are
candidate YSOs in our selection scheme and both of which appear to be
good spectroscopic YSO candidates.

In both $\lambda$ Ori and $\sigma$ Ori a significant fraction (28/75
and 21/48 respectively) of spectroscopically observed K14 {\it WISE}
YSOs do not show significant H$\alpha$ or \ion{Ca}{2} 8542~\AA\ in
emission. However, above $J-K_S$ = 1.3, a majority of {\it WISE}
selected YSOs do exhibit emission in these
lines. Figure~\ref{fig:color-color} shows that in {\it WISE}
color-space, a similar increase in the prevalence of H$\alpha$ and/or
\ion{Ca}{2} 8542 in emission is seen when $w1-w2>0.3$ and $w2-w3>2.2$.

\begin{figure}
  \centering
  \includegraphics[width=9cm]{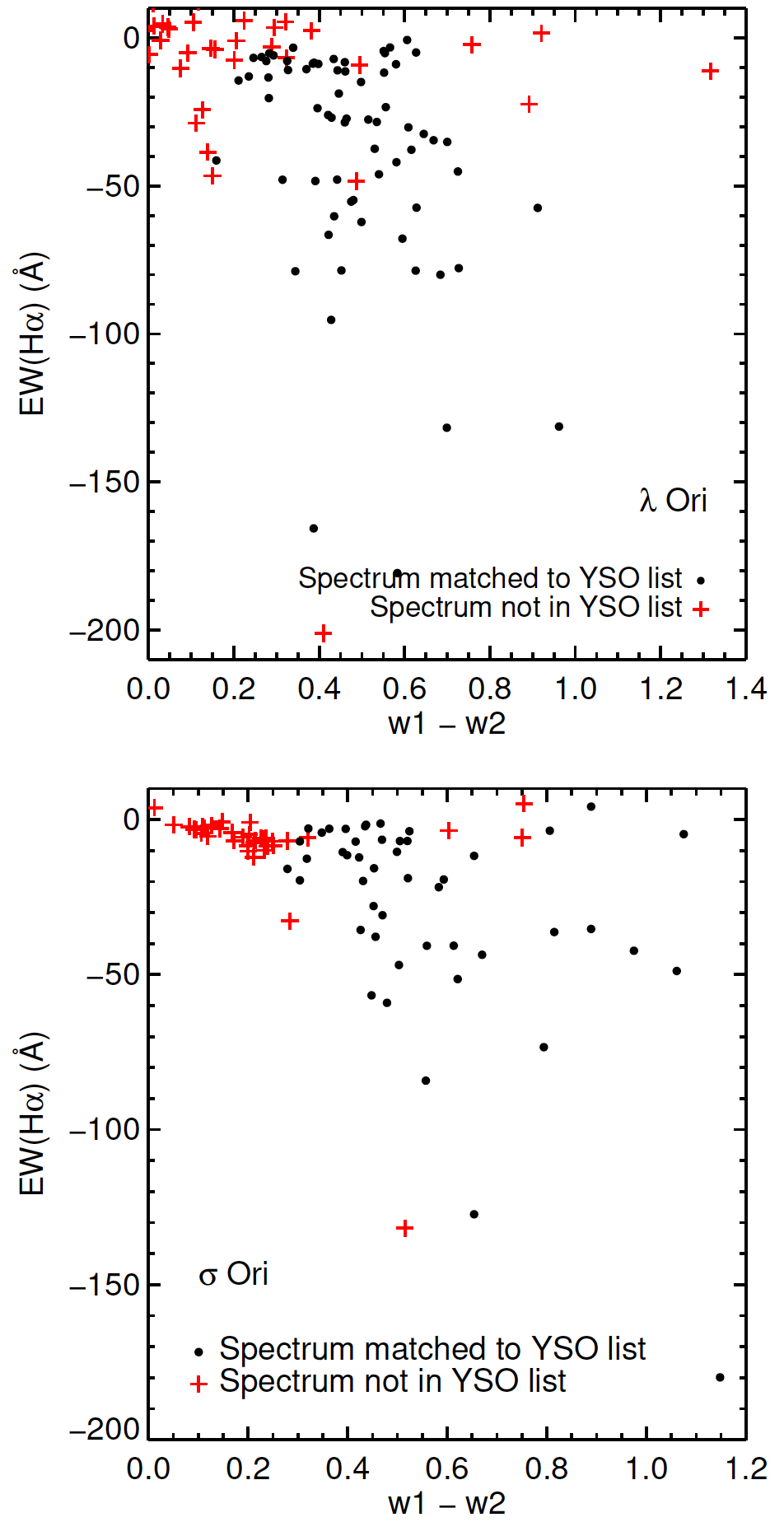}
  \caption{Equivalent width in the H$\alpha$ line versus
    $w1-w2$ color for the spectroscopic sample in the two fields,
    flagged as a YSO candidate by the K14 scheme (black points) or not
    (red points).~\label{fig:halpha12}}
\end{figure}

Objects followed up with spectra also exhibit differing distributions
of H$\alpha$ equivalent width between the K14 YSO candidates and those
spectra not identified by these criteria (i.e., acquired under either
the K12 criteria for infrared excess or as part of the \citet{cody14}
sample). Larger EW(H$\alpha$) in emission (i.e. more negative) are
seen in the K14 selected YSO candidates. In Figure~\ref{fig:halpha12}
we plot the equivalent width in the H$\alpha$ line versus observed
{\it WISE} $w1-w2$ color for the full spectroscopic sample, whether
matched with the K14 YSO catalog (black points) or not (red
crosses). In both fields a K-S test showed a probability $<10^{-4}$
that the equivalent width distributions were drawn from the same
parent distribution. The anomalous red cross points with $w1-w2>$0.6
were not selected as YSOs by the K14 scheme either because they failed
to meet the requisite photometric quality criteria, or were rejected
as AGB candidates because they have very bright $w1$ apparent
magnitude.

We conclude that the K14 YSO selection scheme has been tuned relative
to the K12 YSO selection scheme to be an efficient method of
identifying Class II and Class I objects having significant excess in
2MASS or {\it WISE} filters. The K14 selected YSOs are likely to have
greater amounts of circumstellar material and therefore perhaps higher
accretion rates than those objects with only weak infrared excess.

\subsection{Spectra of Objects Outside the K14 {\it WISE} Selection}

Approximately half of the optical spectra are of objects not
designated as YSO candidates by the K14 scheme.  Specifically, in
$\lambda$ Ori, 55 of the 130 and in $\sigma$ Ori, 52 of 100 objects
for which we acquired spectra do not correspond to a K14 {\it WISE}
YSO candidate.  As described above, this feature is partly because
some of our spectroscopic targets were selected with earlier versions
of our YSO selection criteria and partly because some came from the
separately selected targets of Cody et~al. The K14 scheme aggessively
selects against objects likely to have fake detections in the {\it
  WISE} catalog, in particular in {\it WISE} band 4, which hinders its
ability to find objects that have infrared excess only in the longest
wavelength {\it WISE} band.

As seen in Figure~\ref{fig:halpha12}, at color $w1-w2 < 0.2$ there is
a dearth of infrared-selected YSO candidates.  Yet some of these
objects are moderate H$\alpha$ emitters.  At color $w1-w2 > 0.2$ there
are a large number of infrared-selected YSO candidates, but still a
few objects not selected that are moderate H$\alpha$ emitters, and
conversely some objects that are selected by K14 but do not exhibit
strong H$\alpha$ emission (those with small equivalent widths).

In Table~\ref{tab:missed} we present a summary of the properties of
objects not selected under K14 but for which we have spectra. We list
first the total number of spectra that do not match to a candidate
{\it WISE} YSO, then the number that exhibit H$\alpha$ or \ion{Ca}{2}
8542 in emission above our threshold equivalent width values
($N(young)$). We then give the number that have red infrared colors,
either $w2-w3 > 1$, $w2-w4 > 2$ or excess color $H-K_S > -1.76 \times
(w1-w2) + 0.9$ ($N(excess)$).
Some of these objects not selected under the K14 scheme have weak or
no excess in {\it WISE} bands 1 and 2: $W1-W2 < 0.25$ which makes them
candidate transition disk objects, that is, stars with cleared out or
depleted inner regions. The number of these is given in
Table~\ref{tab:missed} as $N(TD-like)$. Of the full sample of spectra
that do not match K14 {\it WISE} YSO candidates, 24/107 (22\%) show
spectroscopic signs of youth in H$\alpha$ or \ion{Ca}{2} 8542.

\input{tab2.tex}

\subsection{Contaminants: Spatial Distribution and Colors}\label{sec-contam}

Spectra marked in Figure~\ref{fig:color-color} that appear to be those
of AGNs or quasars in the $\lambda$ Ori field are a result of the
overlap in {\it WISE} color space of these extragalactic sources with
protostars and the reddest Class II sources. These objects escape
elimination because of their brightness in $w1$.

\begin{figure}
  \centering
  \includegraphics[width=9.5cm]{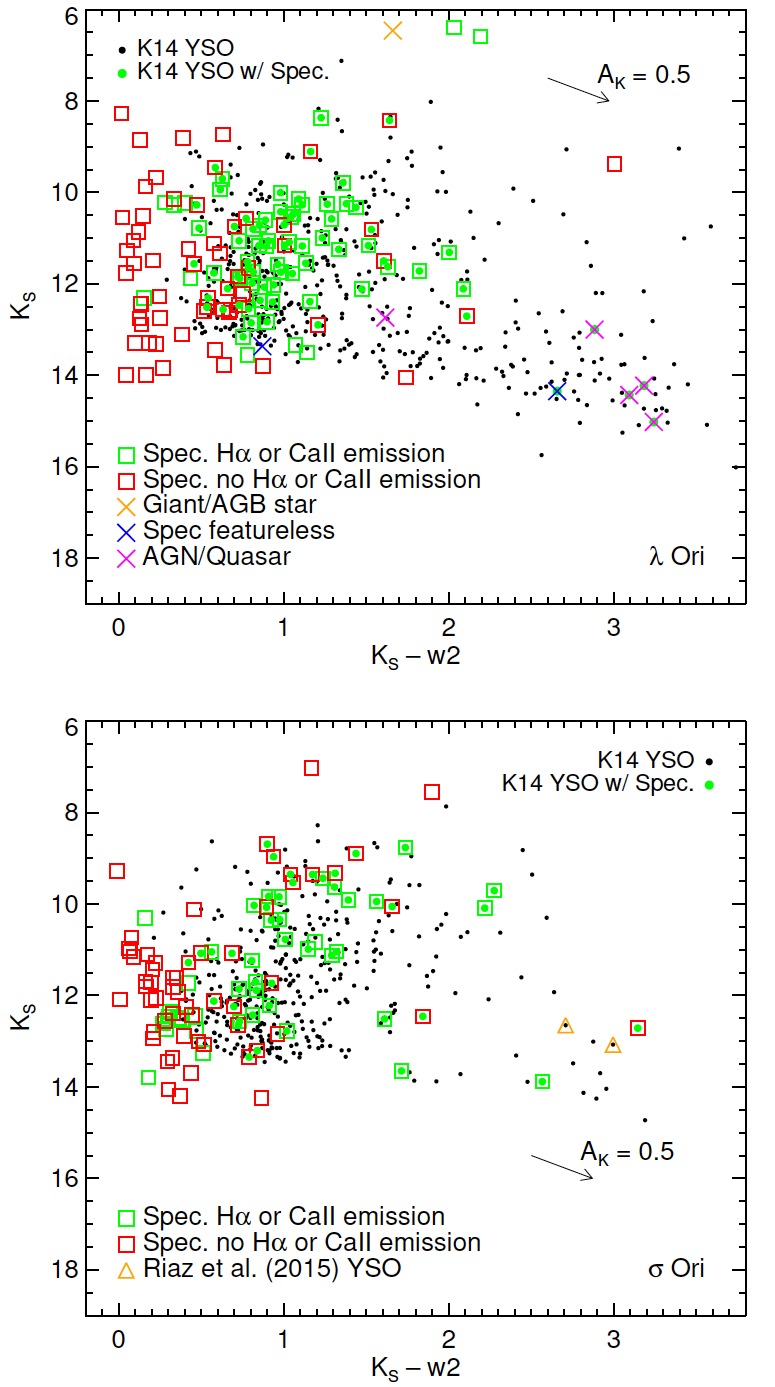}
  \caption{Infrared color-magnitude diagrams of the K14-selected YSO
    candidates in $\sigma$ and $\lambda$ Orionis (black points, or
    green points if a spectrum was acquired). Additional plot symbols
    show spectroscopic targets, colored as in figure
    legend.~\label{fig:kk2}}
\end{figure}

In the $\lambda$ Orionis field, of the K14 YSO candidates with
spectral follow-up, four were found to resemble those of extragalactic
objects, whether AGNs or other galaxies. One K14 Class I candidate
object was found to be a known quasar in a SIMBAD cross-match. One
object appeared to have a featureless but blue spectrum. In
Figure~\ref{fig:kk2} we show a color-magnitude diagram summarizing the
results of the overlap between the {\it WISE} photometric YSO
selection scheme and our spectral follow-up program. In the panel for
$\sigma$ Ori, we also show the location of the two YSOs studied by
\citet{riaz}.

\begin{figure}
  \centering
  \includegraphics[width=12cm]{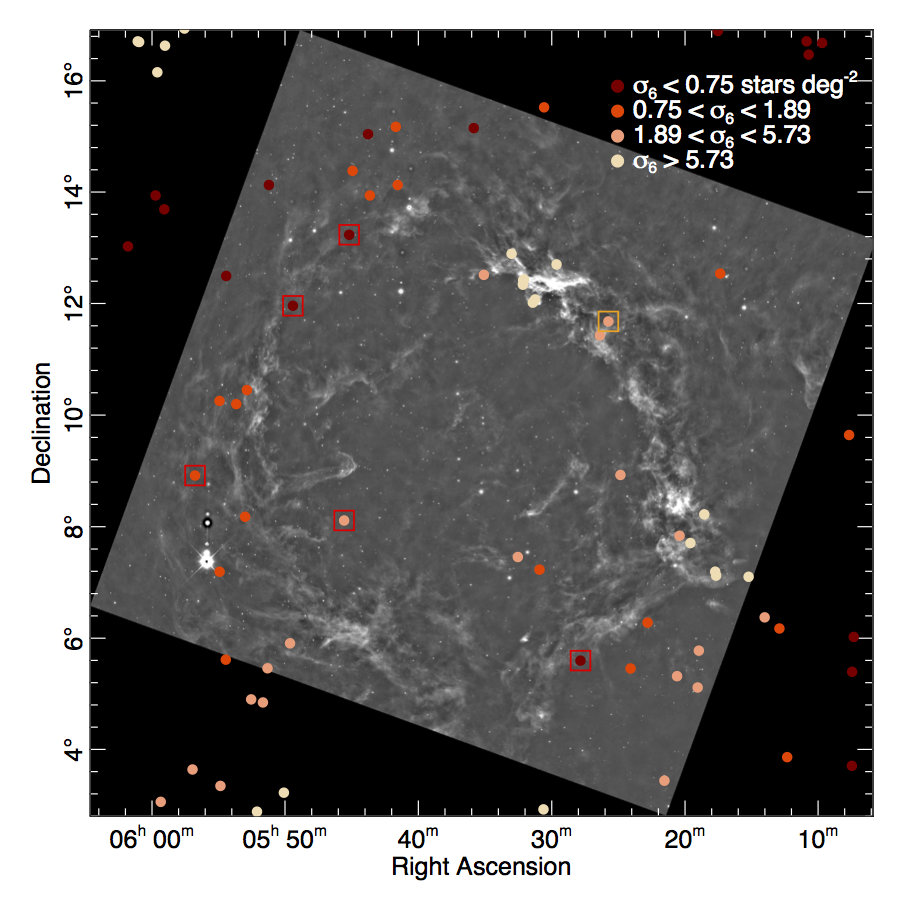}
  \caption{Distribution of protostar candidates in the extended
    $\lambda$ Orionis field color-coded by their local apparent space
    density. Red boxes: AGNs or quasars, orange box: spectra blue and
    featureless.~\label{fig:agn}}
\end{figure}

The extragalactic spectra were found amongst the protostar candidates
in the $\lambda$ Orionis field and in general were found in regions of
low YSO surface density. In Figure~\ref{fig:agn} we show the
distribution of protostar candidate sources color-coded by their
projected angular space density $\sigma_6$, divided into quartiles of
apparent space density. We compute $\sigma_6$ by finding the distance
to each object's 6th nearest neighbor in the complete YSO catalog
(Class II, transition disk, as well as protostars) and following the
prescription of \citet{casert85}. The quartiles are as follows---the
lowest 25\% of protostars have $\sigma_6<$0.75 stars per square
degree, the 2nd quartile 0.75$<\sigma_6<$1.89 stars per square degree,
3rd quartile 1.89$<\sigma_6<$5.73 stars per square degree and the 4th
quartile have $\sigma_6>$5.73 stars per square degree, color-coded
from dark to light red. In the same figure we overlay the location of
the objects whose spectra (or SIMBAD types) were extragalactic or
otherwise non-stellar.

\begin{figure}
  \centering
  \includegraphics[width=16cm]{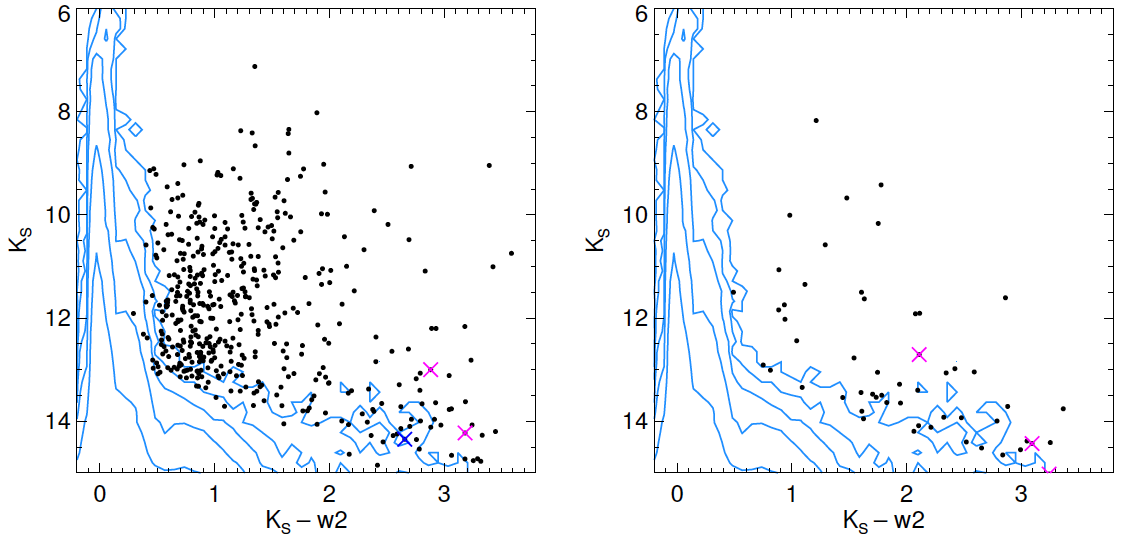}
  \caption{Color-magnitude diagrams of all $\lambda$ Ori {\it WISE}
    YSO candidates above (left panel) and below (right panel)
    apparent space density $\sigma_6$=0.75 stars per square
    degree. The background contours shows a 2D histogram of objects
    drawn from an equivalent area of sky around the North and South
    Galactic poles in the AllWISE catalog (200 square
    degrees). Contour levels log-spaced starting at 1, increasing in
    intervals of 0.7dex. Colored points are as in Fig.~\ref{fig:kk2}
    upper panel.~\label{fig:kk2hess}}
\end{figure}

The colors of low density YSO candidates also show they may be largely
dominated by extragalactic contaminants. In Figure~\ref{fig:kk2hess}
we show color-magnitude diagrams of all {\it WISE} YSO candidates
above and below apparent density $\sigma_6$=0.75 stars per square
degree. We also show for comparison a contour representation of the
color-magnitude distribution of stars within $\sim10 \degr$ of the
North and South Galactic poles (requiring $w2$ signal to noise $>$ 5,
$\chi^2_2<1.8$ and $K_S$ uncertainty $<0.1$). The plot of YSO
candidates at low spatial density shows a relative increase in the
quantity of sources with $K_S - w3 > $4 and $K_S>$12.5, similar to the
region in color-magnitude space seen in the Galactic pole reference
field data. This location is also where the objects found in the
Galactic pole reference fields are found. This trend fits the
expectation that the majority of stars are born in clusters, while
extragalactic contaminants should be found in a more uniform
distribution.

\citet{koenig14} estimated a contamination rate by extragalactic
sources by randomly, repeatedly placing an appropriate number density
of AllWISE catalog objects, drawn from a patch of sky away from the
Galactic Plane, behind the 2D extinction map of \citet{schlegel} for
their Outer Galaxy test field and finding the average number of these
that would be selected as YSO candidates by their scheme. Using the
numbers from their Table~12 and scaling for the area of our Orion
fields, we can estimate that 41 of the 77 Class I candidates and 32 of
the 452 Class II candidates in $\lambda$ Ori may be extragalactic
contaminants. In $\sigma$ Ori the numbers are 4 of 24 and 3 of 381
respectively. K14 estimated the Galactic contamination rate using the
Galactic stellar populations model of \citet{wains} to predict an
upper limit to the number density of non-YSOs that would appear in
their test field in the same infrared magnitude and color ranges as
YSOs in their scheme. We estimate that 28 of the Class I candidates
and 93 of the Class II candidates in $\lambda$ Ori may be Galactic
contaminants and in $\sigma$ Ori, 3 Class I and 9 Class II's
respectively may be Galactic contaminants. Our spectroscopic sample in
$\lambda$ Ori confirms none of the 6 Class I sources observed and 40
of the 64 Class II's as young stars on the basis of H$\alpha$ or
\ion{Ca}{2} 8542 emission line strength. Our spectroscopic sample in
$\sigma$ Ori (together with the samples of Cody et al. and Riaz et
al.) confirms 3 of the 4 Class I sources observed and 26 of the 46
Class II's as young stars in the same way. The Class I success rates
rate and K14 predicted contamination rates appear to be consistent for
the two fields. For the Class II sources, we confirm only the smaller
subset of our candidates that possess H$\alpha$ or \ion{Ca}{2} 8542 in
emission, and miss out on those objects that may have Lithium in
absorption. A more precise test of the contamination rate predictions
would require a measurement of the Lithium absorption line strength
and a larger spectroscopic sample.

The analysis of \citet{koenig14} showed that the YSO candidate lists
generated by their selection scheme were likely contaminated at a rate
of roughly 173 extragalactic sources and at most 290 Galactic sources
in their $\sim$480 square degree test field in the Outer Galaxy, or
0.36~sources deg$^{-2}$ and 0.6~deg$^{-2}$ respectively. These
estimates predict 72 extragalactic and 121 Galactic sources
contaminating the $\lambda$ Orionis {\it WISE} YSO list and 7 and 12
sources in the $\sigma$ Orionis list. The number of extragalactic
contaminants was estimated in K14 by randomly, repeatedly placing an
appropriate number density of AllWISE catalog objects, drawn from a
patch of sky away from the Galactic Plane, behind the 2D extinction
map of \citet{schlegel} and finding the average number of these that
would be selected as YSO candidates by the K14 scheme. The Galactic
contamination rate was estimated by using the Galactic stellar
populations model of \citet{wains} to predict an upper limit to number
density of non-YSOs that would appear in that part of the sky in the
same infrared magnitude and color ranges as YSOs in the K14
scheme. Using the spectroscopic data we have for the two fields
combined (including the samples of Cody et al. and Riaz et al.), for
those sources selected by K14 as Class II YSOs, we find 66 out of 110
objects followed up with spectra show either H$\alpha$ or \ion{Ca}{2}
8542 in emission. For the Class I sources, 3 out 10 objects show
either H$\alpha$ or \ion{Ca}{2} 8542 in emission and we find 6 likely
extragalactic contaminants, whether AGNs, quasars or featureless
spectra. The total predicted contamination rate using the estimation
of K14 above is $\sim$22\% for the two fields combined. This rate
appears low for the Class I sources, albeit from a small sample. The
success rate of confirmed Class II candidates in these fields appears
low as well; however, the absence of lithium absorption equivalent
width measurements for these objects means there may be more young
stars within this sample than we can ascertain from emission lines
alone.

We note that \citet{assef} quote a true AGN (dusty quasar) surface
density of 62 per square degree, based on the simple selection
criteria $w1-w2 \geq 0.8$ and $w2 < 15.05$ in high latitude
fields. This density of objects is much higher than the general
density of all YSO candidates in the two Orion fields in this
paper. In $\sigma$ Ori the median local space density is
33~deg$^{-2}$, while in $\lambda$ Ori it is 9.5~deg$^{-2}$. The AGN
surface density of \citet{assef} is also higher than the predictions
of \citet{koenig14} for the extragalactic contamination rate. K14
estimated that their YSO candidate list was likely contaminated by
approximately 173 extragalactic sources in their $\sim$480 square
degree test field in the Outer Galaxy, or 0.36~sources deg$^{-2}$. The
high surface density of AGNs found by \citet{assef} is largely due to
their deeper survey in $w2$. Our YSO candidate sample is entirely
brighter than $w2=13.5$, resulting from a cut in $w1$ introduced in
the K14 selection scheme specifically to try to minimize the number of
AGNs included in the YSO candidate sample. The extragalactic objects
in our spectroscopic sample are simply the bright tail of the galaxy
population. In general, a $w1$, $w2$ or $K$ magnitude upper limit is a
critical criterion for avoiding extragalactic objects in a
spectroscopic survey.

\subsection{Ancillary Information on Candidates}
Tables~\ref{tab:sig} and \ref{tab:lam} present the entire {\it
  WISE}-2MASS derived catalogs of YSO candidates for the $\sigma$ and
$\lambda$~Orionis fields respectively. We give the AllWISE catalog
designation, coordinates, 2MASS and {\it WISE} photometry, source YSO
class determined by the K14 scheme and the computed SED slope
$\alpha$. We list the equivalent width measurements in the H$\alpha$
and \ion{Ca}{2} 8542~\AA\ lines and spectral types where available
from our spectra.

As a final step in vetting the list of mid-infrared selected YSO
candidates, we consulted Palomar DSS and SDSS images to determine
whether sources could be readily identified as galaxies, and also
SIMBAD\footnote{http://simbad.u-strasbg.fr} for previous information
in the literature regarding possible status as a contaminant or a
young star. Based on this information we have designated sources in
Tables~\ref{tab:sig} and \ref{tab:lam} with a "y" (yes) or "m" (maybe)
or "n" (no) to indicate the likelihood of membership in the Orion star
forming region. Where spectral types were available in the literature
we have included these in the tables and noted the specific reference
in the final column.

In $\lambda$ Ori, of 545 K14-selected candidates, 349 have no
additional membership information from either the literature or our
spectra. Of the 196 with supplemental information beyond just the
$2MASS+WISE$ photometry, we designate 149 as probable members
(including 50 from our spectra) and 26 (including 17 from our spectra)
as possible members for a confirmation rate of 76--89\%. In $\sigma$
Ori, of 418 K14-selected candidates, 147 have no additional membership
information from either the literature or our spectra.  Of the 271
with supplemental information, 233 are probable members (including 38
from our spectra) and 33 (including 6 from our spectra) are possible
members, resulting in a confirmation rate of 86--98\%.  Only 20
sources in $\lambda$ Ori and 4 in $\sigma$ Ori are ruled out as
galaxies or evolved stars.

\section{Final YSO Census}

\subsection{Completeness}
A summary of the number of YSO candidates found using the K14 scheme
is presented in Table~\ref{tab:k14class}. Assuming a 400~pc distance,
the $K_s < 14$ limit of the selected YSO candidates illustrated in
Figure~\ref{fig:kk2} corresponds at an age of 1~Myr to a mass of
0.05~$M_\odot$, with the mass limit increasing as the stars get older
(about 0.15~$M_\odot$ at 10~Myr and 0.5~$M_\odot$ at 1~Gyr). In the
presence of even modest extinction, we should be sensitive to the vast
majority of young star candidates having infrared excess.

To assess the completeness of the K14-derived YSO sample, we take as a
comparison sample the updated list of young stars found in nearby star
forming regions from the Cores to Disks (c2d) survey produced by
\citet{hsieh}. We take their {\it Spitzer} photometric catalog,
classify their young stars with the scheme of \citet{gutermu09} and
produce lists of Class I, II and transition disk young stars. We then
search for that subset of objects in the AllWISE catalog and classify
them following K14. At an apparent magnitude of $K_S$=13.25 (10.3) the
retrieval rate of {\it Spitzer} Class I or II YSOs by {\it WISE} is
down to 50\% (90\%). An apparent magnitude of $K_S$=13.25 converts to
a mass limit of 0.1~M$_\sun$ (1.4~M$_\sun$) for an age between
3-5~Myr, using the stellar evolution models of \citet{siess}, assuming
a distance of 400~pc and zero extinction.

\subsection{YSO Spatial Distribution}

In Figure~\ref{yso-dist}, we show the distribution on the sky of YSOs
found using the scheme of \citet{koenig14}. Figure~\ref{spec-dist}
shows the distribution of the spectra we acquired for this paper.

We note in Figure~\ref{fig:kk2hess} that not all of the low density
YSOs are contaminants. Several YSO candidates in the right panel have
$K_S < 12.5$ and 0.8$<K_S - w2 < $2, a locus that is not a feature of
the reference field color-magnitude distribution. These are likely
young stars that either have drifted away from their birth clusters or
that formed in relative isolation.

We assess the clustering properties of the YSO candidates in Orion by
constructing a minimal spanning tree from the source distribution. An
MST is defined as the network of lines, or branches, that connects a
set of points together such that the total length of the branches is
minimized and there are no closed loops. We follow the methodology of
\citet{gutermu09} to identify a branch length cut off that separates
groups or clusters from the low density background.

\begin{figure}
  \centering
  \includegraphics[width=16cm]{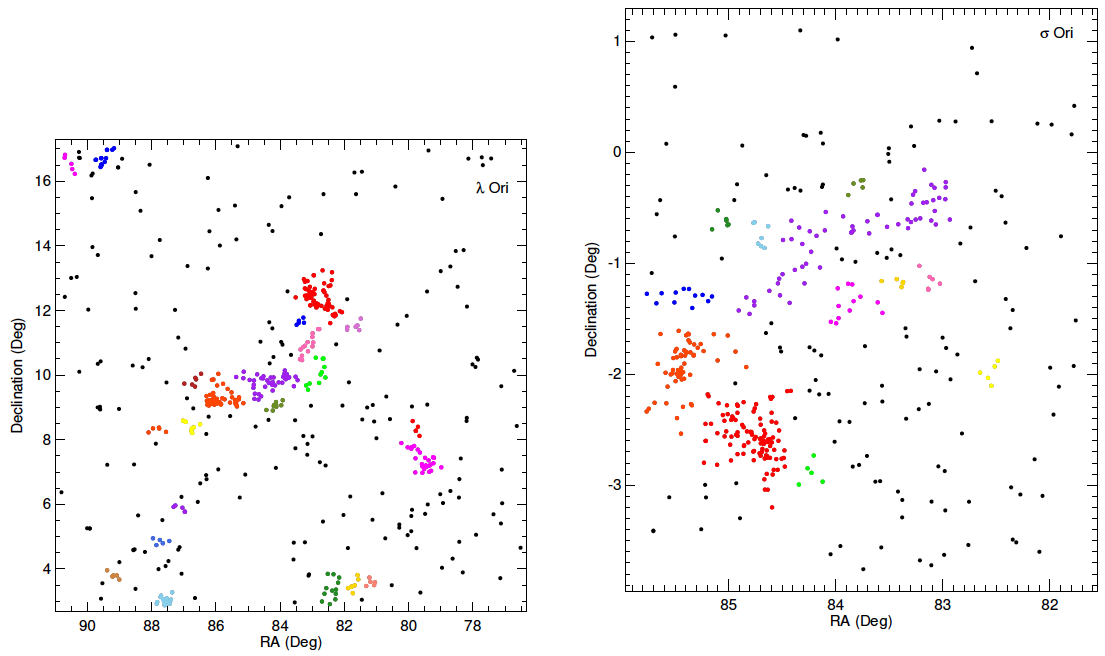}
  \caption{MST groups (colored points) and all other YSO candidates
    (black points) in $\sigma$ and $\lambda$ Ori, left and right
    panels respectively. Stars are colored according to their group
    for visual identification.~\label{fig:mst}}
\end{figure}

We fit two line segments to the cumulative distribution function of
MST branch lengths: a steep-sloped segment at short spacings and a
shallow-sloped segment at long spacings. We then adopt the MST branch
length of the intersection point between the two lines as the critical
cutting length. For the purposes of this analysis, we define a cluster
as a group of five or more points all with branch lengths shorter than
the cutoff. We run this analysis on both regions, using the complete
YSO candidate lists and show the resulting clusters in
Figure~\ref{fig:mst}. In $\lambda$ Ori, 65\% of the YSO candidates
belong to MST-identified groups of 5 or more members, while in
$\sigma$ Ori the fraction is 69\%.

In the right panel, showing the $\lambda$ Ori region, the MST analysis
picks out several stellar aggregates away from the $\lambda$ Ori
ring. To the extreme north-east (RA $>89\degr$, Dec $>16\degr$), it
identifies a YSO cluster associated with the more distant G192.16
\htwo\ region. Along the southern part of the panel, small groups near
to L1617 are seen (RA $>87\degr$, Dec $<6\degr$), as well as the
northern edge of the larger 25 Ori/Orion OB1a cluster (81 $<$ RA
$<$83$\degr$, Dec $<4\degr$). Within the $\lambda$ Orionis ring our
catalog of YSO candidates resides in clusters around the head of the
B35 pillar (85 $<$ RA $<88\degr$, 8$<$Dec $<10\fdg5$) and around B30
(RA $\sim82\fdg$7, Dec $\sim12\fdg$5). In between these groups the MST
algorithm divides the YSOs into small groups around the central
$\lambda$ Ori cluster (purple, pink and green groups). The majority of
the ring of emission seen in {\it WISE} band 3 does not appear to
possess a large quantity of young stars with infrared excess, however
small clusters of Class II and I sources are coincident with L1588 and
L1589 (magenta and red points, 79 $<$ RA $<81\fdg5$, 5$<$Dec
$<9\degr$).

In the right panel showing the $\sigma$ Ori region, the $\sigma$ Ori
cluster itself is identified by the MST algorithm (red points,
centered on RA $\sim84\fdg$7, Dec $\sim -2\fdg$7), as are the groups
associated with NGC~2024 and 2023 (orange points, RA$>85\fdg2$,
$-2.6<$Dec $\sim -1\fdg$5). The more distributed population of YSOs to
the north and west of these regions (83$<$RA $<85\degr$, $-2.6<$Dec
$<$ $-0\fdg$2) is also picked out, but broken up by the MST algorithm
into several groups. We are uncertain about the nature of the
remaining, small groups in the right panel of the figure without
further spectroscopic surveys to probe to lower masses and find and
young stars without disk excess emission.

Using our spectroscopic sample, we confirm 12 YSO candidates outside
MST-defined groups in $\lambda$ Ori and 1 outside the MST-defined
groups in $\sigma$ Ori. The lowest local stellar density around such
objects, as characterized by $\sigma_6$ (see
Section~\ref{sec-contam}), is 0.5 stars per square degree in $\lambda$
Ori and 18.4 stars per square degree in $\sigma$ Ori. This diffuse
population of YSOs could be made up of stars that have dispersed from
where they formed. As with the remainder of the young stars we
identify though, a complete census of the mass function and the
non-disked population is necessary to determine the exact nature of
these objects.

\section{Discussion}

\subsection{The Wide-Field distribution of YSOs in $\sigma$ and $\lambda$ Ori}
While {\it WISE} lacks the sensitivity and resolution to match the
depth of previous YSO surveys in Orion,
Figures~\ref{yso-dist},~\ref{lit-dist} and~\ref{spec-dist} demonstrate
its ability to capture the basic distribution of star formation on
large areal scales.

\subsubsection{$\sigma$ Ori region}
In and around $\sigma$ Orionis itself, several extensive and deep
surveys of YSOs have focused on mapping out the low mass end of the
initial mass function, resulting in the large number of literature
sources marked in Fig.~\ref{lit-dist}. In the brightest region of
nebular emission in the field, the {\it Spitzer} survey of
\citet{megeath} also identifies a much greater number of YSO
candidates than our present {\it WISE} survey, roughly 370 YSOs
deg$^{-2}$, compared to 84 YSOs deg$^{-2}$ from {\it WISE} in that
location, or a factor $\sim$4 greater. In the remaining northern and
western parts of the region, the spectroscopic study of
\citet{briceno} and our YSO candidates comprise a more diffuse
population of YSOs. As can be seen in Figure~\ref{yso-dist}, the {\it
  WISE} YSO candidates are mostly Class II or transition disk
objects. The mismatch between the literature YSOs and our sample in
this part of the field is due to the ability of Brice{\~n}o et~al. to
find weak-line T Tauri stars (WTTs) that lack significant infrared
excess which our color-criteria cannot easily pick out. We find only 2
of the 105 objects noted by Brice{\~n}o et~al. as WTT objects in this
region, but 21 of their 33 `CTT' objects in a 2$\arcsec$ cross-match
between catalogs. As discussed by Brice{\~n}o et~al., these northern
YSOs are a part of the Orion OB1b association. We do not note an
obvious preferential association of our YSO candidates in this area
with the cloud boundary/the bright rim traced by $w3$ emission
however.

\subsubsection{$\lambda$ Ori region}
In $\lambda$ Ori our catalog of YSO candidates traces the previously
observed distribution of YSOs that extends from the head of the B35
pillar across to B30, passing through the central $\lambda$ Ori
cluster. Our catalog captures a subset of the YSO distribution mapped
out by the {\it Spitzer} survey of \citet{hernandez10} around the
central $\lambda$ Ori cluster. We also miss a fraction of the YSOs
documented by the works of Dolan \& Mathieu that lack strong infrared
excess. The small cluster associated with L1588 and L1589 appears to
be new to this study. Previous work in this area has focused on the
several Herbig-Haro objects and outflows (HH114, 115, 328 and 329)
concentrated around IRAS 05155+0707 \citep[see for
  example][]{connell}. Our catalog also finds a low density
distribution of candidate Class I sources to the east and north-east
of $\lambda$ Orionis, but still within the greyscale mosaic area in
Figure~\ref{yso-dist}. Two of these objects have already been found to
be background AGNs (see Section~\ref{sec-contam}). As noted by
\citet{koenig14}, the {\it WISE} Class I candidate selection is known
to be the most highly contaminated by extragalactic sources. Because
of their lack of obvious association with stellar groupings or bright
nebular emission, we suspect that most of this group are likely
background contaminants.

\section{Conclusions}

We have conducted a sensitive search down to the hydrogen burning
limit for unextincted stars using {\it WISE} over $\sim$200 square
degrees around Lambda Orionis and 20 square degrees around Sigma
Orionis. We used the methodology of \citet{koenig14} that builds on
the heritage of the work of \citet{koenig12}, to identify a sample of
544 stars in Lambda and 418 stars in Sigma that are candidate YSOs.

We conducted optical spectroscopic followup of 14\% of these
candidates in Lambda and 11\% in Sigma Orionis. On the basis of strong
emission in the H$\alpha$ or \ion{Ca}{2} 8542 \AA\ lines, we confirm
41 and 27 young stars respectively in the two regions. More reliable
spectroscopic data covering the \ion{Li}{1} 6707 \AA\ line would
enable us to confirm or reject a greater number of the {\it WISE}
selected YSO candidates. Based on our followup spectroscopy for some
candidates and the existing literature for others, we found that
$\sim$80\% of the K14-selected candidates are probable or likely
members of the Orion star forming region.

We improved our understanding of the yield from the photometric
selection criteria in various ranges of color-color and
color-magnitude space, in particular {\it WISE} sources with $K_S -w3
> 1.5$ mag and $K_S $ between 10--12 mag were most likely to show
spectroscopic signs of youth. {\it WISE} sources with $K_S -w3 > 4$
mag and $K_S > 12$ were often AGNs when followed up spectroscopically.

While we improved the census of young stars in the $\lambda$ and
$\sigma$ Ori regions, the presence of several strong \ion{Ca}{2}
and/or H$\alpha$ emitters among spectra obtained of stars not selected
via the K14 methods suggests that the criteria are---though robust and
reliable---not 100\% complete in finding the young stars that are
present.

The population of newly identified and candidate YSOs roughly traces
the known areas of active star formation, but we identify a few new
`hot spots' of activity near Lynds 1588 and 1589. We confirm the more
dispersed population of YSOs seen in the northern half of the
\htwo\ region bubble around $\sigma$ and $\epsilon$ Ori surveyed in
\citet{briceno}. External to the $\lambda$ Orionis ring we note
clusters of YSO candidates around the L1617 cloud and the more distant
G192.16$-$3.82 \htwo\ region.

We present a minimal spanning tree analysis of the two regions to
identify stellar groupings. We find that roughly two-thirds of the YSO
candidates in each region belong to groups of 5 or more members. Given
the likely rate of contamination of the YSO candidate sample by
galaxies and older stars, as described in Section~\ref{sec-contam}, we
suspect the fraction of YSOs in clusters to be higher than this
value. The population of stars selected by {\it WISE} outside the MST
groupings does contain spectroscopically verified YSOs however, with a
local stellar density as low as 0.5 stars per square degree.

\input{stub.tab3.tex}

\input{stub.tab4.tex}

\clearpage

\end{document}

%% file: tab1.tex
\begin{deluxetable}{cccccc}
\tablewidth{0pt} 
\tabletypesize{\scriptsize}
\tablecaption{$WISE$ YSO Candidate Breakdown}
\tablehead{\colhead{Method} & \colhead{N(Class I)} & \colhead{N(Flat)} & \colhead{N(Class II)} & \colhead{N(Class III)} & \colhead{N(Transition Disk)}}
\startdata
 & \multicolumn{5}{c}{$\lambda$ Ori} \\
K12 & 405 & \nodata & 1621 & \nodata & 12653 \\
K14 & 77 & \nodata & 452 & \nodata & 15 \\
$\alpha$ & 58 & 80 & 353 & 53 & \nodata \\
 & \multicolumn{5}{c}{$\sigma$ Ori} \\
K12 & 159 & \nodata & 854 & \nodata & 2855 \\
K14 & 24 & \nodata & 381 & \nodata & 13 \\
$\alpha$ & 9 & 27 & 300 & 82 & \nodata
\label{tab:k14class}
\enddata \tablecomments{K12 refers to the scheme of Koenig
  et~al. (2012), specifically as described in Section 3. K14 refers to
  the scheme of Koenig \& Leisawitz (2014). $\alpha$ refers to YSO
  classes based on the slope of the infrared SED.}
\end{deluxetable}

%% file: tab2.tex
\begin{deluxetable}{ccccc}
\tablewidth{0pt} 
\tabletypesize{\scriptsize}
\tablecaption{Spectroscopic targets not selected by {\it WISE} scheme}
\tablehead{\colhead{Field} & \colhead{Total} & \colhead{N(Young)} & \colhead{N(excess)} & \colhead{N(TD-like)}}
\startdata
$\lambda$ Ori & 55 & 10 & 35 & 22 \\
$\sigma$ Ori & 52 & 15 & 6 & 0
\label{tab:missed}
\enddata
\tablecomments{TD-like means `transition-disk' like, as discussed in the text.}
\end{deluxetable}

%% file: stub.tab3.tex
\begin{deluxetable}{cccccccccccccccccccccccc}
\rotate
\tablewidth{0pt} 
\tabletypesize{\tiny}
\tablecaption{$\sigma$ Orionis $WISE$ YSO Candidates}
\tablehead{\colhead{WISE name} & \multicolumn{2}{c}{Coordinates} & \colhead{J} & \colhead{uJ} & \colhead{H} & \colhead{uH} & \colhead{K} & \colhead{uK} & \colhead{w1} & \colhead{uw1} & \colhead{w2} & \colhead{uw2} & \colhead{w3} & \colhead{uw3} & \colhead{uw4} & \colhead{w4} & \colhead{Class} & \colhead{$\alpha$} & \colhead{EW(H$\alpha$)} & \colhead{EW(\ion{Ca}{2}) 8542} & \colhead{Member} & \colhead{Source Type} & \colhead{Reference} \\ \colhead{ } & \colhead{RA (deg)} & \colhead{Decl. (deg)} & \colhead{(mag)} & \colhead{(mag)} & \colhead{(mag)} & \colhead{(mag)} & \colhead{(mag)} & \colhead{(mag)} & \colhead{(mag)} & \colhead{(mag)} & \colhead{(mag)} & \colhead{(mag)} & \colhead{(mag)} & \colhead{(mag)} & \colhead{(mag)} & \colhead{(mag)} & \colhead{ } & \colhead{ } & \colhead{\AA\ } & \colhead{\AA\ } & \colhead{ } & \colhead{ } & \colhead{ }}
\startdata
J052701.95$-$013053.3 & 81.758142 & $-$1.514806 & 10.143 & 0.023 & 9.587 & 0.022 & 9.295 & 0.023 & 8.836 & 0.023 & 8.515 & 0.020 & 6.379 & 0.016 & 4.243 & 0.029 & II & $-$0.90 & \nodata & \nodata & \nodata & \nodata & \nodata \\
J052705.47+002507.6 & 81.772798 & 0.418795 & 9.523 & 0.027 & 9.132 & 0.023 & 8.630 & 0.021 & 8.461 & 0.023 & 8.065 & 0.020 & 3.810 & 0.014 & 1.109 & 0.012 & II & $-$0.24 & \nodata & \nodata & y & HAeBe;B9 & HD 290409;Vieira et~al. (2003) \\
J052706.54$-$015530.7 & 81.777261 & $-$1.925207 & 13.562 & 0.022 & 12.930 & 0.022 & 12.567 & 0.024 & 12.252 & 0.023 & 11.877 & 0.022 & 10.490 & 0.091 & 8.202 & 0.266 & II & $-$1.75 & \nodata & \nodata & \nodata & \nodata & \nodata \\
J052711.25+000941.8 & 81.796915 & 0.161632 & 14.110 & 0.070 & 13.520 & 0.084 & 13.197 & 0.077 & 12.815 & 0.041 & 12.556 & 0.042 & 10.340 & 0.099 & 8.089 & 0.425 & II & $-$1.32 & \nodata & \nodata & \nodata & \nodata & \nodata \\
J052735.17$-$004523.6 & 81.896577 & $-$0.756557 & 11.278 & 0.022 & 10.444 & 0.021 & 9.942 & 0.021 & 9.081 & 0.023 & 8.678 & 0.020 & 7.066 & 0.016 & 4.567 & 0.029 & II & $-$0.78 & \nodata & \nodata & y & K5 & Brice{\~n}o et~al. (2005) 
\enddata
\tablecomments{This Table is published in its entirety in the electronic edition of the {\it Astrophysical Journal}.~\label{tab:sig} A portion is shown here for guidance regarding its form and content. Right ascension and Declination coordinates are J2000.0.}
\end{deluxetable}

%% file: stub.tab4.tex
\begin{deluxetable}{cccccccccccccccccccccccc}
\rotate
\tablewidth{0pt} 
\tabletypesize{\tiny}
\tablecaption{$\lambda$ Orionis $WISE$ YSO Candidates}
\tablehead{\colhead{WISE name} & \multicolumn{2}{c}{Coordinates} & \colhead{J} & \colhead{uJ} & \colhead{H} & \colhead{uH} & \colhead{K} & \colhead{uK} & \colhead{w1} & \colhead{uw1} & \colhead{w2} & \colhead{uw2} & \colhead{w3} & \colhead{uw3} & \colhead{uw4} & \colhead{w4} & \colhead{Class} & \colhead{$\alpha$} & \colhead{EW(H$\alpha$)} & \colhead{EW(\ion{Ca}{2}) 8542} & \colhead{Member} & \colhead{Source Type} & \colhead{Reference} \\ \colhead{ } & \colhead{RA (deg)} & \colhead{Decl. (deg)} & \colhead{(mag)} & \colhead{(mag)} & \colhead{(mag)} & \colhead{(mag)} & \colhead{(mag)} & \colhead{(mag)} & \colhead{(mag)} & \colhead{(mag)} & \colhead{(mag)} & \colhead{(mag)} & \colhead{(mag)} & \colhead{(mag)} & \colhead{(mag)} & \colhead{(mag)} & \colhead{ } & \colhead{ } & \colhead{\AA\ } & \colhead{\AA\ } & \colhead{ } & \colhead{ } & \colhead{ }}
\startdata
J050603.60+043923.0 & 76.515035 & 4.656403 & 15.733 & 0.115 & 15.141 & 0.132 & 14.412 & 0.116 & 12.783 & 0.027 & 11.160 & 0.023 & 7.062 & 0.017 & 4.279 & 0.028 & I & 1.08 & \nodata & \nodata & \nodata & \nodata & \nodata \\
J050631.77+082532.7 & 76.632415 & 8.425772 & 13.407 & 0.025 & 12.794 & 0.033 & 12.439 & 0.027 & 12.054 & 0.024 & 11.398 & 0.022 & 9.501 & 0.041 & 8.494 & 0.427 & II & -1.27 & \nodata & \nodata & \nodata & \nodata & \nodata \\
J050651.72+100756.5 & 76.715537 & 10.132371 & 13.440 & 0.022 & 13.089 & 0.031 & 13.010 & 0.029 & 12.714 & 0.024 & 12.198 & 0.024 & 9.557 & 0.045 & 7.324 & 0.126 & II & -0.66 & \nodata & \nodata & \nodata & \nodata & \nodata \\
J050821.22+060202.0 & 77.088449 & 6.033907 & 15.511 & 0.077 & 14.644 & 0.080 & 13.922 & 0.062 & 12.677 & 0.024 & 11.599 & 0.022 & 7.918 & 0.021 & 5.122 & 0.034 & I & 0.56 & \nodata & \nodata & \nodata & \nodata & \nodata \\
J050823.77+070431.2 & 77.099072 & 7.075350 & 14.944 & 0.050 & 14.299 & 0.056 & 13.535 & 0.047 & 12.500 & 0.025 & 11.801 & 0.022 & 9.089 & 0.036 & 6.256 & 0.064 & II & -0.04 & \nodata & \nodata & \nodata & \nodata & \nodata \\
\enddata
\tablecomments{This Table is published in its entirety in the electronic edition of the {\it Astrophysical Journal}.~\label{tab:lam} A portion is shown here for guidance regarding its form and content. Right ascension and Declination coordinates are J2000.0.}
\end{deluxetable}